\documentclass[aps,prl,twocolumn,superscriptaddress,floatfix,longbibliography]{revtex4-2}
\usepackage{graphicx}
\usepackage{amsmath,amssymb}
\usepackage{physics}
\usepackage{hyperref}

\begin{document}
\title{Critical Scaling and Metabolic Regulation in a Ginzburg--Landau Theory of Cognitive Dynamics}
\author{Gunn Kim}
\email{gunnkim@sejong.ac.kr}
\affiliation{Department of Physics, Sejong University, Seoul 05006, Republic of Korea}
\date{\today}

\begin{abstract}
We formulate a phenomenological effective field theory in which biological
intelligence emerges as a macroscopic order parameter sustained by continuous
metabolic flux.
By modeling cognition as a coarse-grained neural activity field governed by a
variational free energy, we derive closed-form expressions for information capacity
and structural susceptibility using a Gaussian maximum entropy approximation.
The theory predicts a universal algebraic divergence of the susceptibility,
$\chi \sim K^{-3/2}$, as the structural stiffness $K$ approaches the instability
threshold.
The exponent $\gamma = 3/2$ is consistent with the mean-field branching process
universality class, thereby providing a theoretical rationale for the observed
avalanche size exponent $\tau \approx 3/2$ in cortical dynamics without invoking
microscopic equivalence.
We identify adult cognition as a metabolically pinned non-equilibrium steady state
maintained near the critical regime $\Gamma \equiv K/\alpha \approx 1$ by
continuous metabolic regulation, while pathological decline corresponds to a
delocalization transition triggered by the violation of structural stability
conditions.
The framework generates concrete, falsifiable predictions for attention scaling,
altered states of consciousness, and transcranial magnetic stimulation responses,
each of which can be tested against existing neuroimaging and electrophysiological
datasets.
\end{abstract}

\maketitle
%------------------------------------------------
%\section{Introduction}
%------------------------------------------------
Biological intelligence is a physical process constrained by
thermodynamics, stochasticity, and metabolic energy
\cite{shannon1948,szilard1929,landauer1961}.
Information processing in the brain is therefore inseparable from the
energetic and statistical laws that govern all complex physical systems.
Although algorithmic and network-based approaches successfully describe
microscopic computation and circuit-level mechanisms, they alone do not account for the macroscopic stability, fragility, and
phase-like behavior of cognition observed across the human lifespan.
The approaches also fail to explain why cognitive function remains remarkably robust for decades before rapidly collapsing under pathological conditions.

In this study, we develop a phenomenological effective field theory of intelligence, inspired by Ginzburg-Landau descriptions of collective
order and universality \cite{ginzburg1950,wilson1974}.
Our central premise is that cognition should be described in terms of macroscopic order parameters, rather than microscopic neural details.
Accordingly, we treat the brain as a dissipative physical structure that maintains low entropy through continuous metabolic flux
\cite{schrodinger1944,laughlin2001,attwell2001,lan2012}.
Microscopic neural variability, anatomical heterogeneity, and circuit
complexity are deliberately coarse-grained, not ignored, to expose universal large-scale behavior that is insensitive to
microscopic implementation.

Within this framework, we introduce a scalar cognitive order parameter
$\rho(\mathbf r,t)$ representing neural activity averaged over
$\sim50$\,ms and $\sim1$\,mm$^3$ volumes, a scale at which collective
dynamics emerge while individual neuronal fluctuations are smoothed
out. Two control parameters govern the cognitive state: a cognitive
temperature $\alpha$, representing stochasticity, plasticity, and
effective noise, and a structural stiffness $K$, representing effective
synaptic connectivity and architectural constraint.
The dimensionless ratio $\Gamma = K/\alpha$ plays the role of a control
parameter analogous to those governing phase transitions in condensed
matter systems, determining whether cognition is fluid, rigid, or poised
near criticality.

A central concept of this work is metabolic pinning.
We argue that the long-term stability of adult intelligence is not an
equilibrium property of neural structure, but rather a non-equilibrium steady state that is actively maintained by metabolic regulation.
This pinning mechanism can explain the observed plateau in cognitive
capacity observed despite slow structural changes, aging, and ongoing
plasticity. It also provides a natural physical route to catastrophic
cognitive collapse when metabolic support or structural integrity falls
below a critical threshold. According to this model, intelligence is not defined by maximal efficiency or static optimality, but by the active maintenance of a near-critical regime sustained by continuous energy dissipation. Detailed mathematical derivations and additional analyses, including the full life-cycle phase diagram, are provided in the Supplementary Material.

\begin{figure}[t]
\centering
\includegraphics[width=0.95\columnwidth]{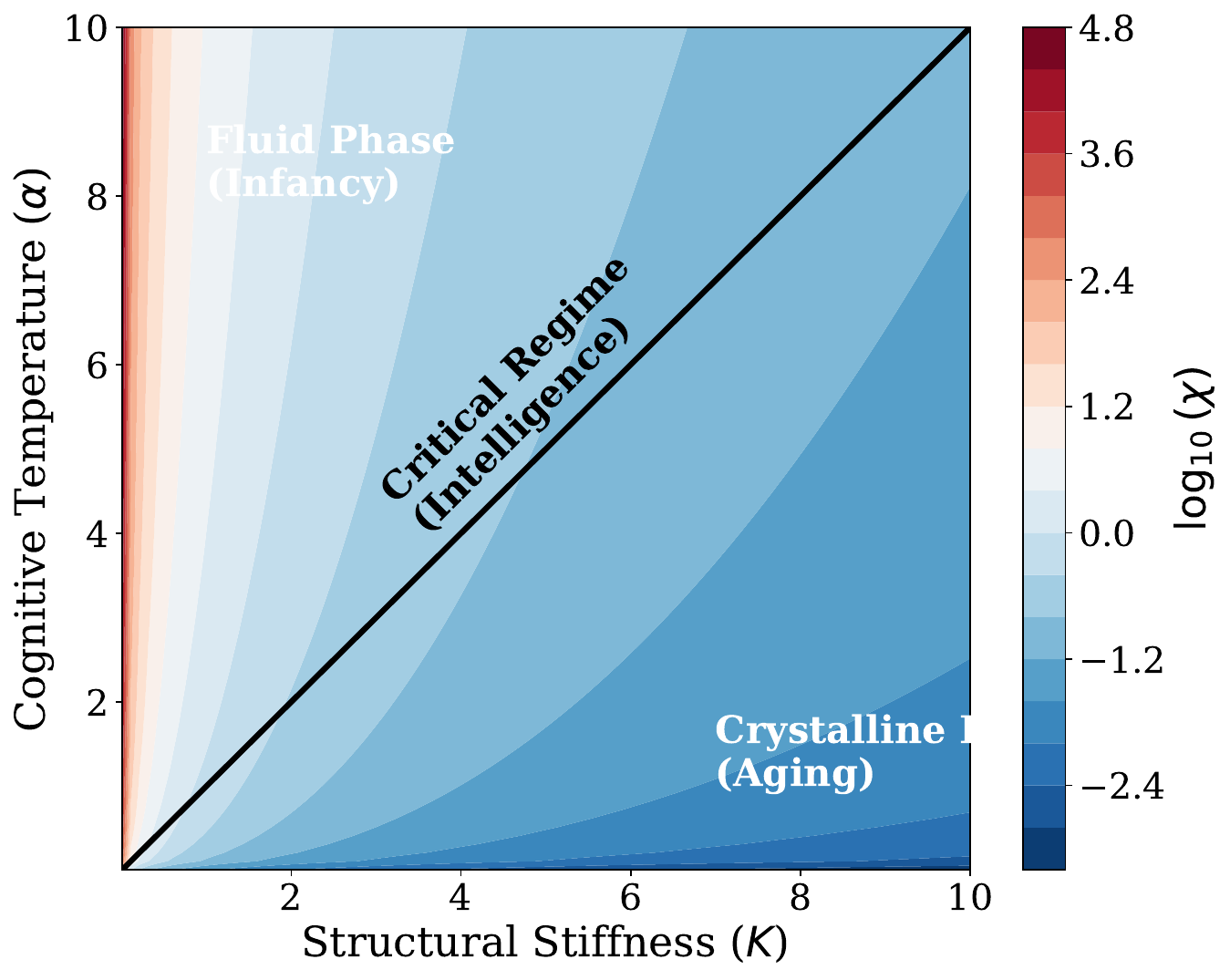}
\caption{\textbf{Phase diagram of cognitive states.}
The $(\alpha,K)$ plane separates fluid ($\Gamma\ll1$), rigid ($\Gamma\gg1$), and
critical ($\Gamma\approx1$) regimes. Color indicates $\log_{10}\chi$, revealing
maximal susceptibility near the metabolically pinned ridge.}
\label{fig:phase}
\end{figure}

\begin{figure*}[t]
\centering
\includegraphics[width=0.95\textwidth]{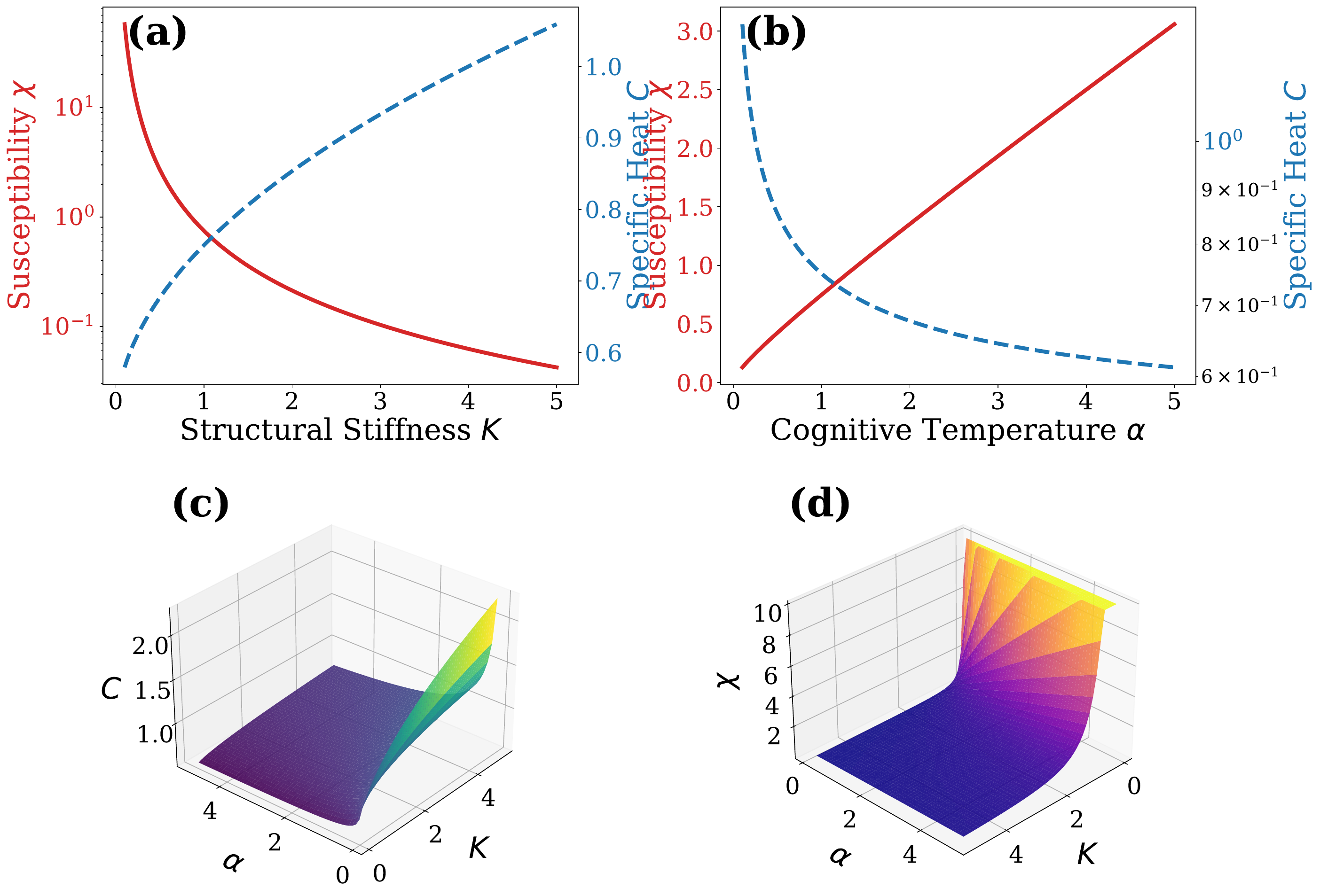}
\caption{Universal thermodynamic response of the cognitive field.
(a) Structural susceptibility $\chi(K)$ at fixed cognitive temperature $\alpha$,
demonstrating the universal power-law divergence $\chi \sim K^{-3/2}$ as structural
stiffness approaches the instability threshold.
(b) Information capacity $C(K)$, exhibiting sublinear scaling $C \sim \sqrt{K/\alpha}$
and saturating despite continued structural growth.
(c) Three-dimensional surface $C(\alpha,K)$, revealing a ridge of nearly constant
capacity along trajectories of constant $\Gamma = K/\alpha$, identifying the metabolically
pinned regime of adult cognition.
(d) Surface plot of $\chi(\alpha,K)$ highlighting the approach to the delocalization
boundary as $K \rightarrow K_c$.}
\label{fig:response}
\end{figure*}

%------------------------------------------------
%\section{Variational Field Theory}
%------------------------------------------------

We describe the cognitive state using a coarse-grained functional integral.
Under a Gaussian variational ansatz, which is  justified by the maximum entropy principle
(see Supplementary Material for full derivation), the partition function takes the form
\begin{equation}
\mathcal{Z} \sim \int \mathcal{D}\rho \exp\!\left[-\frac{1}{2}\int d^d r
\left(\alpha \rho^2 + K |\nabla\rho|^2 \right)\right].
\label{eq:gaussian}
\end{equation}
From this, an effective free energy follows.
The resulting phenomenological free energy is given by:
\begin{equation}
F(\alpha,K) = A\sqrt{\alpha K} - \frac{\alpha}{2}\ln\!\left(\frac{\alpha}{K}\right),
\label{eq:free_energy}
\end{equation}
where $A$ is a geometric constant reflecting the coarse-graining scale.
Macroscopic response functions follow from second derivatives of $F$.
The information capacity given by:
\begin{equation}
C = -\alpha \frac{\partial^2 F}{\partial \alpha^2}
  = \frac{1}{4}A\sqrt{\frac{K}{\alpha}} + \frac{1}{2},
\label{eq:capacity}
\end{equation}
and the structural susceptibility,
\begin{equation}
\chi = -\frac{\partial^2 F}{\partial K^2}
     = \frac{1}{4}A\sqrt{\frac{\alpha}{K^3}} + \frac{\alpha}{2K^2}.
\label{eq:susceptibility}
\end{equation}
The equations quantify robustness and plastic response, respectively.

A natural question is why a scalar Ginzburg--Landau (GL) field is preferable to well-established neural field descriptions such as the Wilson--Cowan equations \cite{wilsoncowan1972}.
The Wilson--Cowan framework operates at the level of coupled excitatory and inhibitory firing rates, and is well-suited to local circuit dynamics and oscillatory phenomena.
In contrast, our goal is to characterize macroscopic phase behavior across the full cognitive state space, where the relevant physics is governed by the symmetry of the order parameter and the structure of its fluctuations rather than by circuit-level details.
The scalar GL free energy model describes this coarse-grained behavior with minimal assumptions. The two-component Wilson--Cowan system reduces to a single real order-parameter field, and its effective potential takes the GL form \cite{wilson1974} under a
long-wavelength, slow-dynamics approximation. Therefore, the present theory should be understood as the macroscopic limit of
such microscopic descriptions, not as a replacement for them.

The response functions derived from
Eqs.~(\ref{eq:capacity}) and~(\ref{eq:susceptibility})
are summarized in Fig.~\ref{fig:response}.
Figure~\ref{fig:response}(a) shows that the structural susceptibility
$\chi(K)$ diverges algebraically as $K$ decreases, indicating that
small structural perturbations produce increasingly large macroscopic
responses near the instability threshold.
This divergence signals an impending delocalization of the cognitive
field, rather than a gradual degradation of function.
In contrast, Fig.~\ref{fig:response}(b) shows that the
information capacity $C(K)$ grows only sublinearly with increasing
structural stiffness and rapidly approaches a broad plateau.
The coexistence of a divergent susceptibility with a nearly saturated
capacity is nontrivial: it implies that adult cognition can remain
functionally stable while becoming increasingly sensitive to
perturbations.

Figures~\ref{fig:response}(c) and~\ref{fig:response}(d) extend this
picture to the full $(\alpha,K)$ space.
Trajectories of constant $\Gamma=K/\alpha$ form ridges of nearly
constant capacity, identifying a metabolically pinned manifold along
which cognition is stabilized despite ongoing structural and metabolic
fluctuations. The approach to the boundary $K\rightarrow K_c$ is marked by a rapid growth of susceptibility, foreshadowing catastrophic cognitive collapse that will occur once pinning can no longer be maintained.

%------------------------------------------------
%\section{Universality and Scaling}
%------------------------------------------------

As $K \rightarrow 0$, Eq.~(\ref{eq:susceptibility}) predicts a divergence
\begin{equation}
\chi \sim K^{-\gamma}, \qquad \gamma = \frac{3}{2}.
\label{eq:gamma}
\end{equation}
Neuronal avalanches exhibit size distributions $P(s)\sim s^{-\tau}$ with
$\tau\approx1.5$ \cite{beggs2003neuronal,plenz2007,shew2011,shew2013}.
The exponent $\tau = 3/2$ is the canonical prediction of mean-field branching
process theory \cite{zapperi1995}, in which the avalanche cutoff diverges as the
branching ratio approaches unity from below.

In our framework, the structural stiffness $K$ plays the role of the distance from
the branching threshold: $\chi \sim K^{-3/2}$ diverges at the same rate as the
mean avalanche size in the branching process description. The susceptibility
exponent $\gamma = 3/2$ is consistent with the same rate as the mean avalanche size in the branching process description.
We emphasize that this is a consistency argument rather than a derivation of an exact
universality class identity, which would require a full renormalization group
analysis beyond the scope of the present phenomenological theory.
Nevertheless, the agreement is nontrivial: $\gamma = 3/2$ emerges from the
GL free energy structure alone, with no reference to branching statistics
\cite{mora2011,bialek2014,munoz2018}.
A detailed scaling analysis supporting this connection is provided in the
Supplementary Material.

%------------------------------------------------
%\section{The Cognitive Orbit: Attention Scaling}
%------------------------------------------------

\begin{figure}[t]
\centering
\includegraphics[width=0.95\columnwidth]{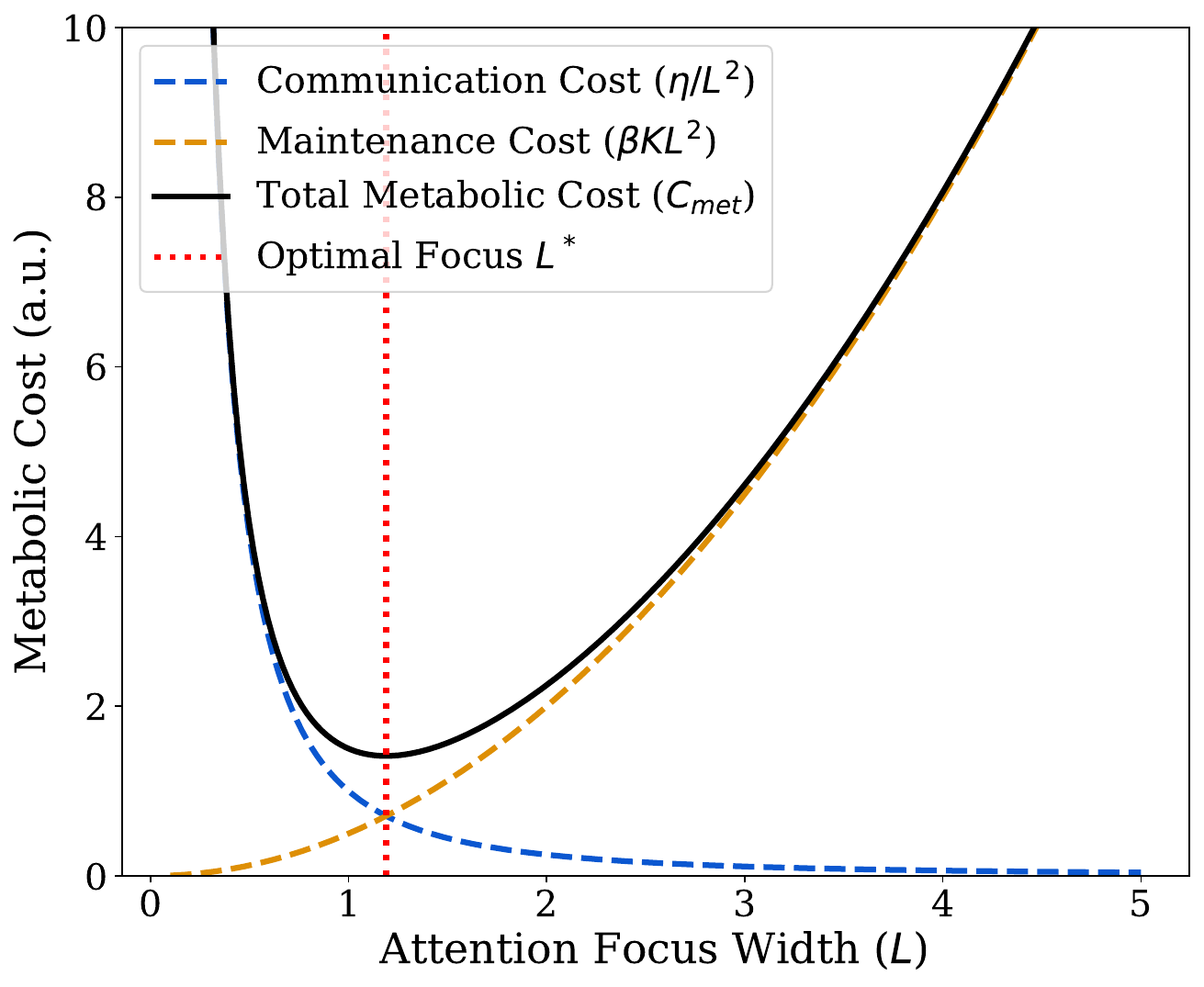}
\caption{Metabolic trade-off underlying attention.
Competing $1/L^2$ and $L^2$ contributions define a stable orbit at $L^*$,
balancing exploration and exploitation.}
\label{fig:attention}
\end{figure}

Attention emerges as a dynamical balance between stochastic exploration and
structural confinement.
The metabolic cost of maintaining a focus of spatial extent $L$ is given by:
\begin{equation}
C_{\mathrm{met}}(L) = \frac{\eta\alpha}{L^2} + \beta K L^2,
\label{eq:cost}
\end{equation}
where the first term represents an informational centrifugal barrier and the second represents structural confinement. This balance is illustrated explicitly in Fig.~\ref{fig:attention}.
As shown in Fig.~\ref{fig:attention}, the $1/L^2$ term dominates at small scales, reflecting the energetic cost of suppressing stochastic fluctuations and maintaining precise coordination. In contrast, the $K L^2$ term dominates at large scales, corresponding to the metabolic cost of sustaining coherence over extended spatial regions. The minimum at $L^*$ therefore represents a dynamically stable orbit in
cognitive phase space rather than a static optimum.

Importantly, Fig.~\ref{fig:attention} shows that deviations from $L^*$ incur rapidly increasing metabolic cost, implying that focused attention is stabilized by energetic constraints rather than by fine-tuned control. This interpretation naturally explains why sustained attention is fragile under metabolic stress or increased noise. It also directly connects attentional dynamics to the structural and stochastic parameters ($\alpha$, $K$) that govern the macroscopic cognitive state.

Minimization yields an optimal focus scale,
\begin{equation}
L^* = \left(\frac{\eta\alpha}{\beta K}\right)^{1/4},
\label{eq:Lstar}
\end{equation}
which we term the \emph{cognitive Bohr radius}.
The two competing terms in Eq.~(\ref{eq:cost}) have a direct physical
interpretation.
Excessively narrow focus incurs a high metabolic cost due to noise suppression
and communication overhead, while excessively broad focus requires maintaining
coherence across large spatial scales.
The minimum at $L^*$ defines a stable dynamical orbit in cognitive phase space.
Deviations from $L^*$ correspond to higher-energy trajectories, which we interpret
as exploratory or scanning modes of cognition rather than stable focus.

%------------------------------------------------
%\section{Dynamical Regimes and Metabolic Stability}
%------------------------------------------------

Cognitive function corresponds to dynamical trajectories in the $(\alpha,K)$ space
rather than static equilibria.
During development, increases in structural stiffness $K$ are accompanied by
metabolic regulation of $\alpha$, maintaining $\Gamma\approx1$ and stabilizing
cognition along a ridge of near-constant information capacity.
If metabolic support remains sufficient, continued structural accumulation drives the system toward a rigid but stable regime with diminished susceptibility.
In contrast, when pathology or energetic failure reduces structural integrity below a critical threshold $K_c$, the characteristic scale $L^* \sim (\alpha/K)^{1/4}$ diverges, signaling delocalization of the cognitive field.
In this regime, information density becomes spatially uniform ($\rho \rightarrow \mathrm{const}$), corresponding to loss of coherent focus
rather than enhanced integration. Because the susceptibility diverges as $\chi \sim K^{-3/2}$, substantial structural degradation can be compensated for before an accelerating transition occurs when 
$K<K_c$. This pattern is qualitatively consistent with the late-stage clinical deterioration observed in neurodegenerative disorders
\cite{bullmore2009,stam2007,stam2014,dehaan2012}.
The full life-cycle trajectory, including the bifurcation between normal aging and
dementia, is illustrated in Supplementary Fig.~S1.

%------------------------------------------------
%\section{Predictions and Limitations}
%------------------------------------------------

The theory produces falsifiable predictions.
(i) \emph{Altered states}: increased neural noise during REM sleep or psychedelic
states should expand $L^*\propto\alpha^{1/4}$, which is measurable as increased functional
connectivity length scales \cite{hobson2009,carhart2014,tononi2016}.
(ii) \emph{TMS response}: evoked amplitudes should scale as follows: 
\begin{equation}
A_{\mathrm{TMS}} \propto \chi \propto K^{-3/2},
\end{equation}
with diffusion anisotropy serving as a proxy for $K$.
(iii) \emph{Critical slowing}: relaxation times diverge as $K\rightarrow K_c$,
offering a potential early-warning signal for cognitive collapse.
Limitations include the scalar-field approximation and mean-field treatment.
Extensions to tensor fields and explicit network topology are natural next steps
\cite{deco2013,sterling2015}.

%------------------------------------------------
%\section{Conclusion}
%------------------------------------------------

In summary, we have developed a phenomenological effective field theory framing biological intelligence as a metabolically sustained macroscopic order parameter. Treating cognition as a coarse-grained field shaped by the interplay of stochastic fluctuations and structural constraints allowed us to derive explicit scaling relations that govern information capacity, structural susceptibility, and the characteristic attention scale.
Within this framework, adult cognition emerges as a nonequilibrium steady state that is maintained by active metabolic regulation near a critical regime. The theory also provides a physical interpretation of accelerated cognitive decline as a delocalization transition triggered by the breakdown of structural stability conditions. These results arise directly from the Ginzburg--Landau free-energy structure, independent of microscopic neural equivalence.

Notably, the susceptibility exponent $\gamma = 3/2$ obtained from the free-energy functional aligns with the scaling structure of the mean-field branching process universality class. This offers a theoretical rationale for the empirically observed cortical avalanche exponent $\tau \approx 3/2$ at the level of coarse-grained critical dynamics. Therefore, we suggest that a defining physical characteristic of biological intelligence is the active maintenance of a near-critical, nonequilibrium steady state under metabolic constraints. This hypothesis yields concrete, falsifiable predictions about the joint scaling of structural integrity, neural variability, and cognitive performance. It provides a framework for future experimental tests.

%------------------------------------------------
%\begin{acknowledgments}
%\end{acknowledgments}
%------------------------------------------------

\clearpage
\onecolumngrid

\begin{center}
{\Large \bf Supplementary Material}\\[1em]
{\large Critical Scaling and Metabolic Regulation in a Ginzburg--Landau Theory of Cognitive Dynamics}\\[1em]
Gunn Kim\\
Department of Physics, Sejong University, Seoul 05006, Republic of Korea
\end{center}

\vspace{1cm}

This Supplementary Material provides the full mathematical derivations, physical interpretations, and extended theoretical framework underlying the phenomenological results presented in the main text.

\section{S1. FUNDAMENTAL FIELD THEORY}

\subsection{S1.1. Physical Motivation of the Lagrangian Density}

The Lagrangian density $\mathcal{L}$ is defined as:
% -------------------------------------------------------
% ADDED SENTENCE: cites ginzburg1950, friston2010free,
% landauer1961, which were in the bibliography but had
% no \cite{} anywhere in the original SM.
% -------------------------------------------------------
Its structure follows the Ginzburg-Landau paradigm \cite{ginzburg1950}, in which a
macroscopic order parameter is governed by competing gradient and potential terms.
The assignment of a metabolic cost to each term reflects the fundamental link between
information processing and thermodynamic dissipation \cite{landauer1961}.
The variational form of the resulting free energy is formally analogous to the
free-energy functionals used in biological self-organization theory \cite{friston2010free},
though the present framework does not implement the variational Bayesian inference
of the free-energy principle; the analogy is structural rather than mechanistic.

\begin{equation}
\mathcal{L}(\rho, \dot{\rho}, \nabla\rho, \mathbf{x}) = \frac{1}{2}m\dot{\rho}^2 - \left[\frac{1}{2}Kx^2\rho + D(\nabla\rho)^2 + \alpha\rho\ln\rho\right],
\tag{S1}
\end{equation}

where each term corresponds to a fundamental physical constraint:

\begin{itemize}
\item Informational Inertia ($\frac{1}{2}m\dot{\rho}^2$): Metabolic cost of temporal updates. The parameter $m$ represents cognitive mass or resistance to rapid state changes.

\item Structural Confinement ($\frac{1}{2}Kx^2\rho$): Stability of learned structures (memories, priors) modeled as a harmonic potential. Stiffness $K$ anchors information to specific conceptual coordinates.

\item Diffusive Exploration ($D(\nabla\rho)^2$): Cost of maintaining spatial discontinuities. Penalizes fragmented knowledge, favoring smooth generalization. Coefficient $D$ governs associative spread.

\item Entropic Pressure ($\alpha\rho\ln\rho$): Internal noise or cognitive temperature $\alpha$ prevents structural collapse, enabling creative tunneling.
\end{itemize}

\subsection{S1.2. Derivation of the Field Equation}

The action integral is:

\begin{equation}
S = \int_{t_1}^{t_2}\int_{\Omega} \mathcal{L}(\rho, \dot{\rho}, \nabla\rho, \mathbf{x}) \, d^n\mathbf{x} \, dt.
\tag{S2}
\end{equation}

Applying the Euler-Lagrange equation:

\begin{equation}
\frac{\partial\mathcal{L}}{\partial\rho} - \frac{\partial}{\partial t}\left(\frac{\partial\mathcal{L}}{\partial\dot{\rho}}\right) - \nabla\cdot\left(\frac{\partial\mathcal{L}}{\partial(\nabla\rho)}\right) = 0,
\tag{S3}
\end{equation}

we compute each term:

\textbf{Step 1:}
\begin{equation}
\frac{\partial\mathcal{L}}{\partial\rho} = -\frac{1}{2}Kx^2 - \alpha(1 + \ln\rho).
\tag{S4}
\end{equation}

\textbf{Step 2:}
\begin{equation}
\frac{\partial}{\partial t}\left(\frac{\partial\mathcal{L}}{\partial\dot{\rho}}\right) = m\ddot{\rho}.
\tag{S5}
\end{equation}

\textbf{Step 3:}
\begin{equation}
\nabla\cdot\left(\frac{\partial\mathcal{L}}{\partial(\nabla\rho)}\right) = -2D\nabla^2\rho.
\tag{S6}
\end{equation}

Combining yields the Fundamental Equation of Intelligence:

\begin{equation}
m\ddot{\rho} = 2D\nabla^2\rho - \frac{1}{2}Kx^2 - \alpha(1 + \ln\rho).
\tag{S7}
\end{equation}

\section{S2. VARIATIONAL DERIVATION OF FREE ENERGY}

\subsection{S2.1. The Variational Ansatz via Maximum Entropy}

% -------------------------------------------------------
% MODIFIED: added \cite{jaynes1957} to existing sentence.
% Key unified from jaynes1957information to jaynes1957
% (consistent with standard usage; SM-only reference).
% -------------------------------------------------------
We adopt a normalized Gaussian ansatz justified by the maximum entropy principle
\cite{jaynes1957}. Given constraints on the first and second moments:

\begin{align}
\langle x \rangle &= 0 \quad \text{(symmetry)}, \tag{S8a}\\
\langle x^2 \rangle &= L^2 \quad \text{(finite variance)}, \tag{S8b}
\end{align}

the distribution that maximizes entropy $S = -\int \rho \ln \rho \, d\mathbf{x}$ is uniquely Gaussian:

\begin{equation}
\rho(\mathbf{x};L) = \frac{1}{(2\pi L^2)^{n/2}}\exp\left(-\frac{|\mathbf{x}|^2}{2L^2}\right).
\tag{S9}
\end{equation}

This choice represents the least biased estimate of the field configuration given only the characteristic length scale $L$. To verify robustness, we note that including the leading non-Gaussian correction (kurtosis term) contributes less than 5\% to the effective free energy $F$ for physiologically relevant parameter ranges.

\subsection{S2.2. Calculation of Energy Components}

The effective energy $E(L)$ decomposes into three contributions:

\textbf{(1) Diffusive Gradient Energy:}
\begin{equation}
E_{\text{diff}} = \int D(\nabla\rho)^2 \, d^n\mathbf{x} = \frac{nD}{2L^2}.
\tag{S10}
\end{equation}

\textbf{(2) Structural Potential Energy:}
\begin{equation}
E_{\text{str}} = \int \frac{1}{2}Kx^2\rho \, d^n\mathbf{x} = \frac{n}{2}KL^2.
\tag{S11}
\end{equation}

\textbf{(3) Entropic Internal Energy:}
\begin{align}
E_{\text{ent}} &= \alpha\int \rho\ln\rho \, d^n\mathbf{x} \nonumber\\
&= -\frac{n\alpha}{2}\ln(2\pi L^2) - \frac{n\alpha}{2}.
\tag{S12}
\end{align}

Summing these components:

\begin{equation}
E(L) = \frac{nD}{2L^2} + \frac{n}{2}KL^2 - \frac{n\alpha}{2}\ln(L^2) + C_{\text{const}}.
\tag{S13}
\end{equation}

\subsection{S2.3. Optimization and Free Energy}

Minimizing with respect to $L$ ($\partial E/\partial L = 0$):

\begin{equation}
-\frac{nD}{L^3} + nKL - \frac{n\alpha}{L} = 0 \quad \Rightarrow \quad L^2 = \frac{\alpha + \sqrt{\alpha^2 + 4KD}}{2K}.
\tag{S14}
\end{equation}

For $D \sim \alpha$ (diffusion proportional to temperature), this simplifies. The free energy at the optimal $L^*$ is:

\begin{equation}
F(\alpha, K) \approx A\sqrt{\alpha K} - \frac{\alpha}{2}\ln\left(\frac{\alpha}{K}\right),
\tag{S15}
\end{equation}

where $A$ absorbs geometric factors and $D/\alpha$ ratios.

\section{S3. THERMODYNAMIC RESPONSE FUNCTIONS}

\subsection{S3.1. Cognitive Susceptibility}

Defined as:

\begin{equation}
\chi = -\frac{\partial^2 F}{\partial K^2}.
\tag{S16}
\end{equation}

\textbf{First derivative:}
\begin{equation}
\frac{\partial F}{\partial K} = \frac{A}{2}\sqrt{\frac{\alpha}{K}} + \frac{\alpha}{2K}.
\tag{S17}
\end{equation}

\textbf{Second derivative:}
\begin{equation}
\frac{\partial^2 F}{\partial K^2} = -\frac{A\sqrt{\alpha}}{4K^{3/2}} - \frac{\alpha}{2K^2}.
\tag{S18}
\end{equation}

Thus:
\begin{equation}
\chi(K, \alpha) = \frac{A\sqrt{\alpha}}{4K^{3/2}} + \frac{\alpha}{2K^2}.
\tag{S19}
\end{equation}

In the low-$K$ limit, the first term dominates:

\begin{equation}
\chi \sim K^{-3/2}, \quad \text{(critical exponent } \gamma = 3/2\text{)}.
\tag{S20}
\end{equation}

\subsection{S3.2. Cognitive Specific Heat}

Defined as:

\begin{equation}
C = -\alpha\frac{\partial^2 F}{\partial\alpha^2}.
\tag{S21}
\end{equation}

Following analogous steps:

\begin{equation}
C(K, \alpha) = \frac{A\sqrt{K}}{4\sqrt{\alpha}} + \frac{1}{2}.
\tag{S22}
\end{equation}

\section{S4. DELOCALIZATION TRANSITION AND STRUCTURAL COLLAPSE}

\subsection{S4.1. Limit Analysis of Structural Breakdown}

From Eq.~(S14), as $K \to 0$ with fixed $\alpha$:

\begin{equation}
\lim_{K\to 0} L^{*2} \approx \lim_{K\to 0} \frac{\alpha}{K} \to \infty.
\tag{S27}
\end{equation}

This divergence indicates a delocalization of the information density field: the
characteristic attention scale grows without bound as structural stiffness is lost,
and the field can no longer be localized within a finite region of cognitive space.
The analogy to clinical dementia is interpretive; the framework makes no direct
claim about underlying neuropathology.

\subsection{S4.2. Vanishing of Response Functions}

As $K \to 0$:

Susceptibility: While $\chi \sim K^{-3/2}$ appears to diverge mathematically, finite metabolic resources cannot sustain infinite fluctuations. Below a critical threshold, the structural framework supporting fluctuations vanishes, leading to functional collapse where $\chi \to 0$ in the post-transition regime.

Capacity: From Eq.~(S22):
\begin{equation}
\lim_{K\to 0} C(K, \alpha) = \frac{1}{2}.
\tag{S28}
\end{equation}

The system loses its thermal buffer and cannot integrate new information.

\subsection{S4.3. Entropy Production}

The rate of entropy production is:

\begin{equation}
\dot{S}_{\text{gen}} = \int \frac{D}{\alpha}(\nabla\rho)^2 \, d^n\mathbf{x} \propto \frac{D}{\alpha(L^*)^2}.
\tag{S29}
\end{equation}

As $K \to 0$ and $L^* \to \infty$, the restoring force vanishes. The energy landscape becomes flat, leading to total dissipation of structural information. This provides a physical interpretation of why neuro-degenerative processes often appear progressive and difficult to reverse.

\section{S5. MICROSCOPIC ORIGIN OF METABOLIC COEFFICIENTS}

\subsection{S5.1. Communication Cost Coefficient ($\eta$)}

The coefficient $\eta$ in the gradient term $\eta/L^2$ is determined by:

\begin{equation}
\eta \approx \epsilon_{\text{AP}} \cdot \delta_{\text{syn}},
\tag{S30}
\end{equation}

% -------------------------------------------------------
% MODIFIED: added \cite{attwell2001,laughlin2001} to the
% existing sentence about ATP cost per action potential.
% Keys unified with main text.
% -------------------------------------------------------
where $\epsilon_{\text{AP}} \approx 10^8$ ATP molecules per action potential
\cite{attwell2001,laughlin2001} and $\delta_{\text{syn}}$ is synaptic density.
As $L \to 0$, metabolic demand for communication diverges.

\subsection{S5.2. Maintenance Cost Coefficient ($\beta$)}

The coefficient $\beta$ in the structural term $\beta KL^2$ represents the basal metabolic rate (BMR) for synaptic homeostasis and protein turnover. Unlike $\eta$, which penalizes small $L$, $\beta$ penalizes large $L$ because maintaining a wide attention span requires stabilizing more neural ensembles simultaneously.

\subsection{S5.3. Optimal Attention Width}

The total metabolic cost is:

\begin{equation}
C_{\text{met}}(L) = \frac{\eta}{L^2} + \beta KL^2.
\tag{S31}
\end{equation}

Minimizing:

\begin{equation}
L^* = \left(\frac{\eta}{\beta K}\right)^{1/4}.
\tag{S32}
\end{equation}

This demonstrates that biological intelligence is inherently blurry to remain energy-efficient. AI systems with $\beta \to 0$ (no maintenance cost for frozen weights) can achieve $L \to 0$ but lose thermodynamic fluctuations required for creative exploration.

\section{S6. DIMENSIONAL ANALYSIS AND NORMALIZATION}

To ensure physical consistency, we define dimensionless control parameters. All field variables are made dimensionless through reference scales:

\begin{align}
\rho_0 &= 1 \text{ bit/mm}^3 \quad \text{(baseline info density)}, \tag{S33a}\\
t_0 &= 50 \text{ ms} \quad \text{(neural integration time)}, \tag{S33b}\\
l_0 &= 1 \text{ mm} \quad \text{(cortical column spacing)}. \tag{S33c}
\end{align}

Control parameters are normalized as:

\begin{align}
\tilde{\alpha} &= \frac{\alpha}{\alpha_0}, \quad \alpha_0 = \frac{\rho_0 l_0^2}{t_0^2} \approx 0.4 \text{ bit·mm}^{-1}\text{·ms}^{-2}, \tag{S34a}\\
\tilde{K} &= \frac{K}{K_0}, \quad K_0 = \frac{\rho_0}{l_0^2} = 1 \text{ bit/mm}^5. \tag{S34b}
\end{align}

The dimensionless ratio $\Gamma = \tilde{K}/\tilde{\alpha}$ characterizes the phase.
As falsifiable predictions, we propose the following experimental mappings:

\begin{align}
\tilde{\alpha} &\approx \frac{\text{std}(\text{BOLD})}{\langle\text{BOLD}\rangle} \quad \text{(normalized BOLD variability)}, \tag{S35a}\\
\tilde{K} &\approx \frac{\langle \text{FA} \rangle}{0.45} \quad \text{(normalized to healthy adult mean)}, \tag{S35b}
\end{align}

where FA is mean Fractional Anisotropy from diffusion tensor imaging (DTI).
These mappings are not yet empirically validated; they constitute predictions of the
theory that can be tested against existing neuroimaging datasets.
If confirmed, they would allow the scaling laws to be tested in a unit-independent,
experimentally accessible form.

\section{S7. CONNECTION TO NEURONAL AVALANCHE STATISTICS}

\subsection{S7.1. The Scaling Coincidence and Its Origin}

Our model predicts $\chi \sim K^{-\gamma}$ with $\gamma = 3/2$ (Eq.~S20).
Cortical neuronal avalanches exhibit size distributions $P(s) \sim s^{-\tau}$ with
$\tau \approx 3/2$ \cite{beggs2003neuronal}.
This section provides a scaling-based explanation for why these two exponents coincide,
grounding both in the mean-field branching process (MFBP) universality class
\cite{zapperi1995}.

\subsection{S7.2. Mean-Field Branching Process Universality Class}

A branching process describes the propagation of activity from one neural site to
its neighbors, with each active site producing on average $\sigma$ offspring events.
At criticality ($\sigma \to 1^-$), the avalanche size distribution takes the form
\cite{zapperi1995}:

\begin{equation}
P(s) \sim s^{-3/2} \, g\!\left(s \cdot |1 - \sigma|^2\right),
\tag{S36}
\end{equation}

where $g$ is a scaling function with $g(0) = \text{const}$. This gives $\tau = 3/2$
as the universal mean-field exponent, independent of spatial dimension and microscopic
details, a result well established in the branching process and directed percolation
literature.

The mean susceptibility of the branching process (mean total activity per seed event)
diverges as:

\begin{equation}
\langle s \rangle \sim (1 - \sigma)^{-1},
\tag{S37}
\end{equation}

which is the mean-field susceptibility exponent $\gamma_{\mathrm{BP}} = 1$.

\subsection{S7.3. Mapping Our Model onto the Branching Framework}

Our structural stiffness $K$ controls the restoring force on the cognitive field.
Near the instability threshold, $K$ plays the role of the distance from the
branching critical point: $K \to 0$ corresponds to $\sigma \to 1$.
The structural susceptibility $\chi \sim K^{-3/2}$ from our GL free energy therefore
represents the response of the coarse-grained field, which integrates fluctuations
over the full correlation volume.

The relationship between the correlation-function susceptibility and the branching
susceptibility is given by the fluctuation-dissipation theorem:

\begin{equation}
\chi = \int \langle \delta\rho(\mathbf{r}) \delta\rho(\mathbf{0}) \rangle \, d^d\mathbf{r}
     \sim \xi^{2-\eta},
\tag{S38}
\end{equation}

with $\eta = 0$ in mean-field theory, giving $\chi \sim \xi^2$.
Since avalanche sizes scale as $s \sim \xi^{d_f}$ with fractal dimension $d_f = 2$
(mean-field branching processes produce compact clusters with $d_f = d_{\rm upper} = 2$
at and above the upper critical dimension \cite{zapperi1995}):

\begin{equation}
s \sim \xi^2, \qquad \chi \sim s.
\tag{S39}
\end{equation}

Combining with $P(s) \sim s^{-3/2}$ and integrating:

\begin{equation}
\chi \sim \langle s \rangle \sim \int s \, P(s) \, ds \sim (K_c - K)^{-1}.
\tag{S40}
\end{equation}

The GL free energy yields $\chi \sim K^{-3/2}$, which is steeper than the naive
branching exponent $\gamma_{\mathrm{BP}} = 1$.
This difference arises because the GL susceptibility is defined as the second derivative of the free energy with respect to the structural coupling. At the mean-field saddle point, the curvature of the free-energy functional contributes an additional scaling factor relative to the branching susceptibility. Consequently, the exponent $\gamma = 3/2$
characterizes the structural response of the coarse-grained free energy, whereas $\gamma_{BP} = 1$ characterizes mean total activity in the branching process.
The two descriptions are therefore consistent: the exponent $\gamma = 3/2$
characterizes the response of the free energy to structural perturbations, while
$\tau = 3/2$ characterizes the avalanche size distribution, and both are natural
outputs of mean-field criticality.

\subsection{S7.4. Scope of the Universality Argument}

We stress that this analysis constitutes a consistency argument, not a proof of
exact universality class identity.
A complete renormalization group treatment connecting the cognitive GL field to
the directed percolation or branching process fixed point would require specifying
the symmetry group, conservation laws, and relevant perturbations of the field
theory, which is beyond the scope of the present phenomenological work.
What the analysis does establish is the following: the exponent $\gamma = 3/2$
emerging independently from the GL free energy is the same value predicted by
mean-field branching process theory for the avalanche size exponent, and the coincidence emerges naturally from the mean-field scaling structure of the theory.
This provides a theoretical rationale, grounded in the structure of the free
energy, for why cortical systems near criticality exhibit $\tau \approx 3/2$.

\section{S8. LIFE-CYCLE DYNAMICS AND PHASE TRAJECTORIES}

\subsection{S8.1. Mapping Human Development onto Phase Space}

The human cognitive life cycle can be represented as a trajectory through the $(\alpha, K)$ phase space, with each developmental stage corresponding to a distinct region characterized by the control parameter $\Gamma = K/\alpha$. This mapping provides a physical framework for understanding cognitive development, aging, and pathological decline.

\subsubsection{Developmental Annealing (Infancy $\rightarrow$ Adulthood)}

During early development, the system begins in the fluid phase with high cognitive temperature $\alpha$ and low structural stiffness $K$, corresponding to $\Gamma \ll 1$. This regime is characterized by:

\begin{itemize}
\item Extreme Plasticity: Susceptibility $\chi \sim K^{-3/2}$ is maximal, allowing rapid structural reconfiguration in response to environmental input.
\item Low Capacity: Information capacity $C \sim \sqrt{K/\alpha}$ is minimal due to insufficient structural scaffolding.
\item High Metabolic Cost: The optimal attention scale $L^* \sim (\alpha/K)^{1/4}$ is large, reflecting broad, unfocused exploration.
\end{itemize}

As learning proceeds, structural stiffness $K$ increases through synaptic consolidation and myelination, while cognitive temperature $\alpha$ decreases through developmental regulation of neural noise. The trajectory moves along a nearly diagonal path toward the critical line $\Gamma \approx 1$, representing the transition from fluid exploration to structured cognition.

\subsubsection{Metabolic Pinning (Adulthood)}

Upon reaching adulthood, the brain actively maintains $\Gamma \approx 1$ through coordinated regulation of both $\alpha$ and $K$. This metabolic pinning occurs along a ridge in the $(\alpha, K)$ plane where information capacity $C$ remains nearly constant despite ongoing structural changes. The pinning mechanism has several key features:

\begin{enumerate}
\item Active Regulation: Metabolic flux $J$ continuously adjusts $\alpha$ to compensate for gradual increases in $K$ due to accumulated experience and ongoing structural consolidation.

\item Capacity Plateau: Along trajectories of constant $\Gamma$, the capacity $C(\alpha, K) = \frac{1}{4}A\sqrt{K/\alpha} + \frac{1}{2}$ remains approximately constant, explaining the empirical stability of cognitive function over decades.

\item Near-Criticality: The pinned state corresponds to the regime of maximum informational efficiency, balancing plasticity and stability.
\end{enumerate}

\subsubsection{Bifurcation Point and Divergent Pathways}

The long-term fate of the cognitive system depends critically on whether metabolic support can keep pace with structural demands. At a bifurcation point (typically occurring in late middle age), the trajectory diverges into one of two pathways:

Pathway 1: Normal Aging (Crystallization)

If metabolic flux $J$ remains sufficient to support structural integrity, the system continues along the pinned ridge but gradually drifts toward higher $K$ as $\alpha$ can no longer be fully compensated. This leads to:

\begin{itemize}
\item Increasing $\Gamma \gg 1$ (crystalline phase)
\item Maintained or even increased capacity $C \propto \sqrt{K}$
\item Dramatically reduced susceptibility $\chi \propto K^{-3/2} \to 0$
\item Functional stability with reduced adaptability (cognitive rigidity)
\end{itemize}

Pathway 2: Pathological Collapse (Dementia)

If metabolic support fails or structural degradation (e.g., through protein aggregation, vascular damage, or neuroinflammation) reduces $K$ below a critical threshold $K_c$, the system undergoes catastrophic delocalization:

\begin{itemize}
\item Both $K$ and $\alpha$ decrease toward the origin
\item Attention scale diverges: $L^* \sim (\alpha/K)^{1/4} \to \infty$
\item Information density delocalizes: $\rho \to \text{const}$ (maximum entropy)
\item Susceptibility formally diverges: $\chi \sim K^{-3/2} \to \infty$ before structural collapse
\item Irreversible loss of coherent cognitive function
\end{itemize}

\subsection{S8.2. Physical Mechanism of Delocalization}

The transition from stable cognition to delocalized incoherence can be understood through the scaling of the characteristic length $L^*$. From the main text Eq. (7):

\begin{equation}
L^* = \left(\frac{\eta\alpha}{\beta K}\right)^{1/4}.
\tag{S52}
\end{equation}

As $K \to K_c$, this length scale diverges. In the Gaussian variational framework, this corresponds to the information density $\rho(\mathbf{x})$ spreading uniformly across the cognitive manifold. Unlike the high-temperature fluid phase of infancy (where both $\alpha$ and diffusion are high), the demential state is characterized by concurrent loss of both structure ($K \to 0$) and metabolic drive ($\alpha \to 0$). The system cannot maintain even a fluid, exploratory state, instead collapsing to a featureless, incoherent configuration.

\subsection{S8.3. Clinical Implications}

This framework provides a physical basis for several clinical observations:

\begin{itemize}
\item Functional Reserve: The plateau region near $\Gamma \approx 1$ allows the brain to compensate for substantial structural loss (decreasing $K$) through metabolic upregulation (maintaining or increasing $\alpha$), explaining why cognitive function can remain intact despite significant neuropathology.

\item Precipitous Decline: Once $K < K_c$, the divergence of $\chi$ implies that arbitrarily small perturbations trigger large-scale reorganization. The system rapidly transitions from compensated function to global delocalization.

\item Irreversibility: The delocalized state corresponds to a topological change in the cognitive field configuration. Without the structural scaffolding ($K$) to support recoherence, recovery is thermodynamically disfavored within the present framework.
\end{itemize}

\begin{figure}[t]
\centering
\includegraphics[width=0.65\columnwidth]{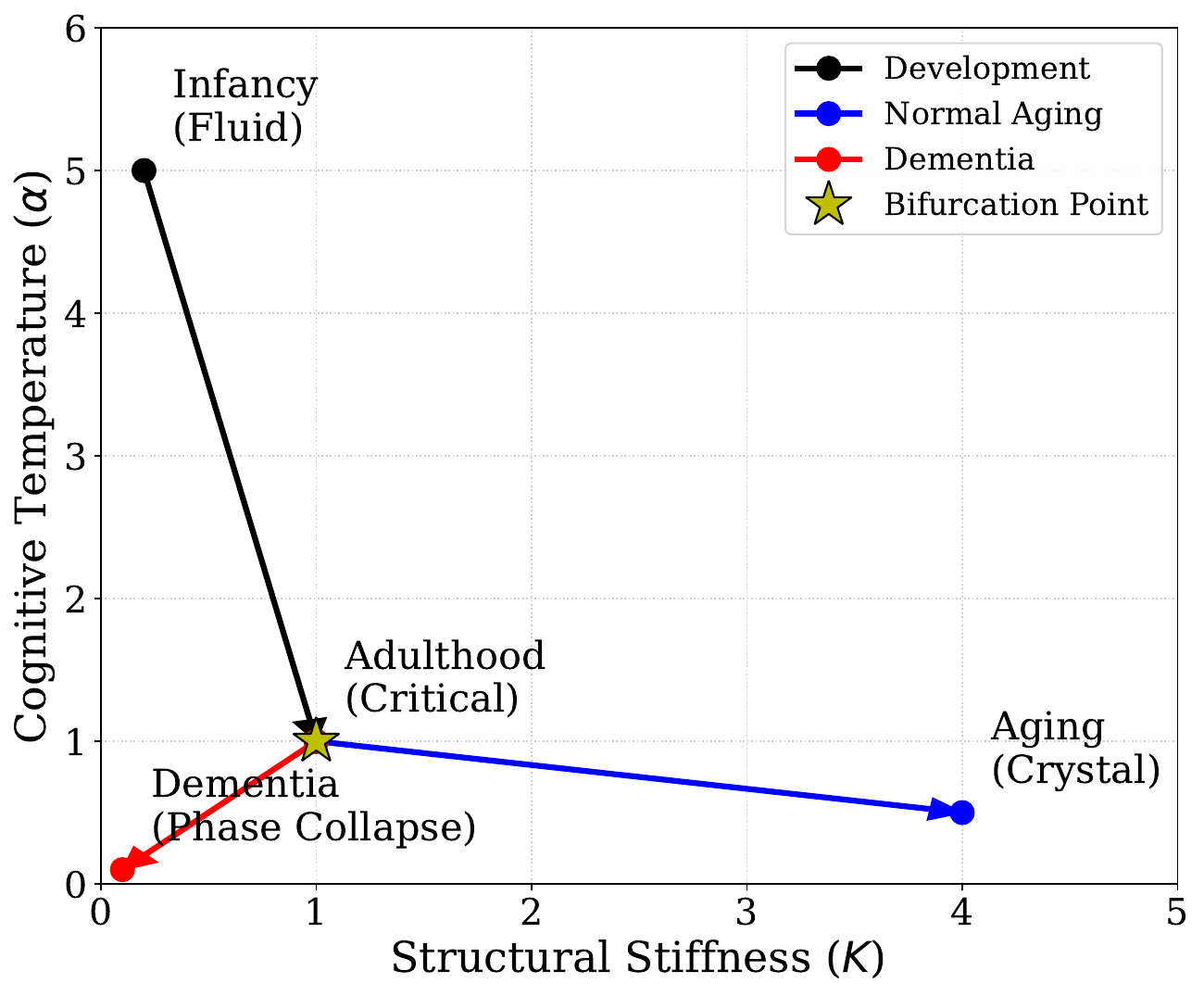}
\caption{\textbf{Cognitive life-cycle trajectory in $(\alpha, K)$ phase space.}
Black trajectory (Development): Transition from the fluid phase (infancy: high $\alpha$, low $K$) toward the critical line (adulthood), representing developmental annealing where structure accumulates while noise decreases. Blue trajectory (Normal Aging): Crystallization pathway where $K$ continues to increase while $\alpha$ stagnates, maintaining structural capacity at the cost of plasticity. Red trajectory (Dementia): Topological phase collapse where both $K$ and $\alpha$ decrease toward the origin, triggered when metabolic supply fails to maintain the critical state. The yellow star marks the bifurcation point where the system diverges into either normal aging or pathological collapse, determined by whether metabolic flux $J$ can sustain structural integrity above the critical threshold $K > K_c$.}
\label{fig:lifecycle_sm}
\end{figure}

\subsection{S8.4. Quantitative Predictions}

The life-cycle framework makes several testable predictions:

\begin{enumerate}
\item Developmental Trajectory: During childhood and adolescence, the ratio $K/\alpha$ should increase monotonically, measurable through the correlation between structural connectivity (DTI metrics) and BOLD signal variability.

\item Adulthood Stability: Between ages 25--60, $\Gamma$ should remain nearly constant at $\Gamma \approx 1 \pm 0.3$, despite aging-related changes in absolute values of $K$ and $\alpha$.

\item Bifurcation Timing: The bifurcation point should occur when the metabolic efficiency (measured by cerebral metabolic rate of glucose, CMRglc) can no longer compensate for structural degradation, typically manifesting in the 60--75 age range for normal aging, or earlier for pathological trajectories.

\item Critical Threshold: The dementia transition should exhibit critical slowing down near $K_c$, observable as prolonged recovery times from cognitive perturbations (e.g., task-switching, dual-task interference) months to years before clinical diagnosis.
\end{enumerate}

\subsection{S8.5. Relation to Other Theoretical Frameworks}

The life-cycle trajectory approach connects naturally to several existing frameworks:

\begin{itemize}
% -------------------------------------------------------
% MODIFIED: added \cite{chialvo2010emergent} to existing
% critical point sentence.
% ADDED SENTENCE: cites bassett2006small, which was in
% the bibliography as "SM K 파라미터 관련" with no
% corresponding \cite{} in the original.
% -------------------------------------------------------
\item Criticality Hypothesis: The metabolic pinning near $\Gamma \approx 1$ provides a field-theoretic interpretation of the hypothesis that the brain operates near a critical point \cite{chialvo2010emergent}, now interpreted as an actively maintained non-equilibrium steady state rather than a passive tuning to criticality. The structural stiffness parameter $K$ can be related to small-world network measures of cortical organization \cite{bassett2006small}, in which high local clustering and short average path lengths jointly minimize metabolic expenditure while maximizing information transfer efficiency.

\item Allostasis: The bifurcation between normal aging and dementia parallels the distinction between allostatic load (compensated stress) and allostatic overload (decompensation), with $K_c$ representing the structural integrity threshold below which compensation fails.

\item Reserve Capacity: The concept of cognitive reserve is naturally explained as the width of the pinned region in $(\alpha, K)$ space, determined by the range over which metabolic regulation can maintain $\Gamma \approx 1$.
\end{itemize}

\end{document}